\newcommand{\be}{\begin{equation}}
\newcommand{\ee}{\end{equation}}

\documentclass[10pt,letterpaper,twocolumn]{article} 

\usepackage{ol2}
\usepackage[draft]{hyperref}
\usepackage{amsmath}

\begin{document}

\twocolumn[ 

\title{Discrete multivortex solitons}

\author{Daniel Leykam and Anton S. Desyatnikov}

\address{Nonlinear Physics Centre, Research School of Physics and Engineering,\\The Australian National University, Canberra ACT 0200, Australia}

\begin{abstract}
We introduce discrete multivortex solitons in a ring of nonlinear oscillators coupled to a central site. Regular clusters of discrete vortices appear as a result of mode collisions and we show that their stability is determined by global symmetries rather than the stability of constituent vortices. Stable multivortex solitons support complex vortex dynamics including charge flipping and spiraling.
\end{abstract}

\ocis{190.5940, 190.6135.}

] 

Discrete vortex solitons have received considerable attention over the past decade~\cite{malomed2001,squares,Falk}, realizing persistent currents between coupled nonlinear oscillators, such as optical and matter-wave waveguides. A stable discrete vortex requires balanced currents between each pair of sites in the loop. For general configurations with multiple vortices the balance condition should include each pair of neighboring vortices and thus it extends over the whole field, making multiple vortices globally linked~\cite{crasovan2002}. Multivortex fields appear naturally in periodic media~\cite{garciamarch2009,JOA2009} but so far the only stable multivortex solitons predicted~\cite{alexander2007} and observed~\cite{terhalle2008} in hexagonal photonic lattices consist of simple alternating vortex-antivortex pairs.

In periodic media the vorticity is restricted by discrete symmetries~\cite{trans,Kartashov} and it is a nontrivial question if stable discrete solitons can be constructed with asymmetric multiple vortices and globally linked currents. The interplay between symmetry and nonlinearity is captured by purely discrete systems~\cite{malomed2001,squares,EilbeckPD}, such as a ring of $N$ coupled oscillators~\cite{PRA11}, governed by the discrete nonlinear Schr\"odinger equation (DNLS). The multivortex solitons can be studied in this simple system if we add a central site~\cite{DC}, similar to hexagonal lattices when $N=6$~\cite{alexander2007,terhalle2008}.

In this Letter, we investigate discrete vortices in a ring of nonlinear oscillators coupled to a central site. We demonstrate that multivortex linear modes and solitons bifurcate at the crossings of modes with different symmetries, controlled by the coupling strength with the central site. We study in detail the heptamer with $N=6$ and find analytically and numerically a variety of stable multivortex solitons. By tracing the trajectories of individual vortices in the stable perturbed solitons we show the appearance of breather-like states with periodic charge flipping and spiraling of individual vortices, i.e. formation of helical vortex structures.

\begin{figure}
\centering\includegraphics[width=1\columnwidth]{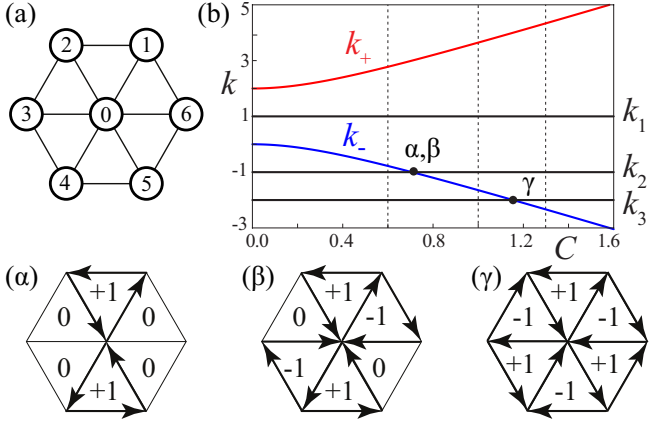}
\caption{(Color online) (a) Heptamer configuration of nonlinear oscillators. (b) Wavenumbers of linear modes vs. central coupling $C$. ($\alpha,\beta,\gamma$) Multivortex modes at crossings shown by dots in (b); the arrows show non-zero currents and the numbers ($-1,0,+1$) indicate topological charges $m_n$ in each triangular cell. The total charge $m=\sum m_n$ is $m=2$ for mode ($\alpha$) and $m=0$ for modes ($\beta,\gamma$).}
\label{fig:array}
\end{figure}

We assume that the sites on the ring with complex amplitudes $E_n$ are identical while the central site $E_0$ can be different, characterized by the coupling strength $C$. The DNLS takes the form
\begin{eqnarray}
i \partial_z E_0 + C \sum E_n +  \delta_0 |E_0|^2 E_0 = 0,\nonumber\\
i \partial_z E_n + E_{n-1} + E_{n+1} + C E_0 +  \delta_1 |E_n|^2 E_n = 0,\label{DNLS}
\end{eqnarray}
here and below the summation is over $n=1,..,N$ sites on the ring, $E_{n+N} = E_n$, and $\delta_0,\,\delta_1 = \mp 1$ stand for de/focusing nonlinearities. This system conserves the total power $P=|E_0|^2+\sum |E_n|^2$, i.e. $dP/dz=0$. The geometry for heptamer with $N=6$ is shown in Fig.~\ref{fig:array}(a).

We look for stationary solutions with propagation constant $k$, $E_0 = A \exp (ikz)$ and $E_n = B_n \exp(i k z)$, and remove the global phase by setting $A \ge 0$,
\begin{eqnarray}
(k - \delta_0 A^2)A &=& C \sum B_n,\nonumber\\
(k - \delta_1 |B_n|^2 ) B_n &=& C A + B_{n-1} + B_{n+1}.
\label{stat}
\end{eqnarray}
In the {\em linear limit} $\delta_{0,1} = 0$ this system is solved with the discrete Fourier transform (DFT), $B_n=\sum_{s=1}^N F_s e^{i\Theta_s n}$, here $\Theta_s= 2\pi s/N$ and $F_s$ are the Fourier amplitudes. First, we find vortex-free (real) modes with $A>0$ and $F_s=0$ for all $s\ne N$,
\be
k_{\pm} = 1 \pm \sqrt{1 + N C^2} = NC F_N/A.
\ee
The ``bright'' mode with $k=k_+$ has all sites oscillating in phase, $F_N>0$, while the ``dark'' mode with $k=k_-$ has the ring sites $\pi$-out-of-phase with the center, $F_N<0$. The rest of the modes are constant-amplitude discrete vortices of charge $|m|\le N/2$ on the ring with zero central site $A=0$~\cite{PRA11}, i.e. with $F_{s\ne m}=0$ and the dispersion relation $k = k_m = 2 \cos \Theta_m$. Figure~\ref{fig:array}(b) shows the dependence of mode propagation constants on the coupling $C$ for a heptamer.

At specific values of the coupling $C$ the $k_-$ curve intersects with $k_m$ lines in Fig.~\ref{fig:array}(b), allowing the dark mode to coexist with vortex modes in a stationary superposition. The collision of two modes creates novel states with distinct topologies, or {\em multivortex modes}, with multiple vortices residing in different elementary triangular cells, as we show schematically in Figs.~\ref{fig:array}($\alpha,\beta,\gamma$). 
While the discrete vortices without the central site, $C=0$, obey the symmetry $E_{n+1} = E_n \exp (i \Theta_m)$, this symmetry is broken for multivortex modes. As a result, $m$ no longer uniquely defines the topology of the mode. For example, in Fig.~\ref{fig:array}(b) the intersection of two modes $k_-$ and $k_2$ supports two different configurations, ($\alpha$) with $m=2$ and ($\beta$) with $m=0$.

We now examine the nonlinear case, starting with {\em constant amplitude solitons}, $B_n = B \exp(i \psi_n)$ and $B > 0$. With this Ansatz the exact solutions can be obtained by applying DFT, $\exp (i \psi_n) = \sum_{s=1}^N b_s \exp(i \Theta_s n)$, thus
\begin{eqnarray}
&&B b_s (k - \delta_1 B^2 - 2 \cos \Theta_s ) = 0, \;\;s \ne N \nonumber \\
&&B b_N (k - \delta_1 B^2 - 2) = C A, \nonumber \\
&&B b_N N C = (k - \delta_0 A^2)A.
\label{solitons}
\end{eqnarray}
As in the linear case above, there are vortex-free solitons with $b_{s\ne N}=0$, as well as the discrete vortex solitons with $A=0$ and $k=k_m=2 \cos \Theta_m +\delta_1 B^2$~\cite{PRA11}.

The {\em discrete multivortex solitons} appear, similar to the linear multivortex modes, as superpositions of a real mode with amplitude $b_N$ and two vortex modes,
\be
e^{i \psi_n} = b_N + b_{+m} e^{i \Theta_m n} + b_{-m} e^{-i \Theta_m n},
\label{epsi}
\ee
with $k=k_m$ and complex amplitudes $b_{\pm m}$ constrained by $|\exp(i \psi_n)| = 1$. We find such solutions for $N/|m| = 2,3,4$ only. In particular, the $N=2|m|$ solitons with alternating charges, as in Fig.~\ref{fig:array}($\gamma$), have been observed experimentally in hexagonal photonic lattices~\cite{terhalle2008}.

\begin{figure}[t]
\centering\includegraphics[width=\columnwidth]{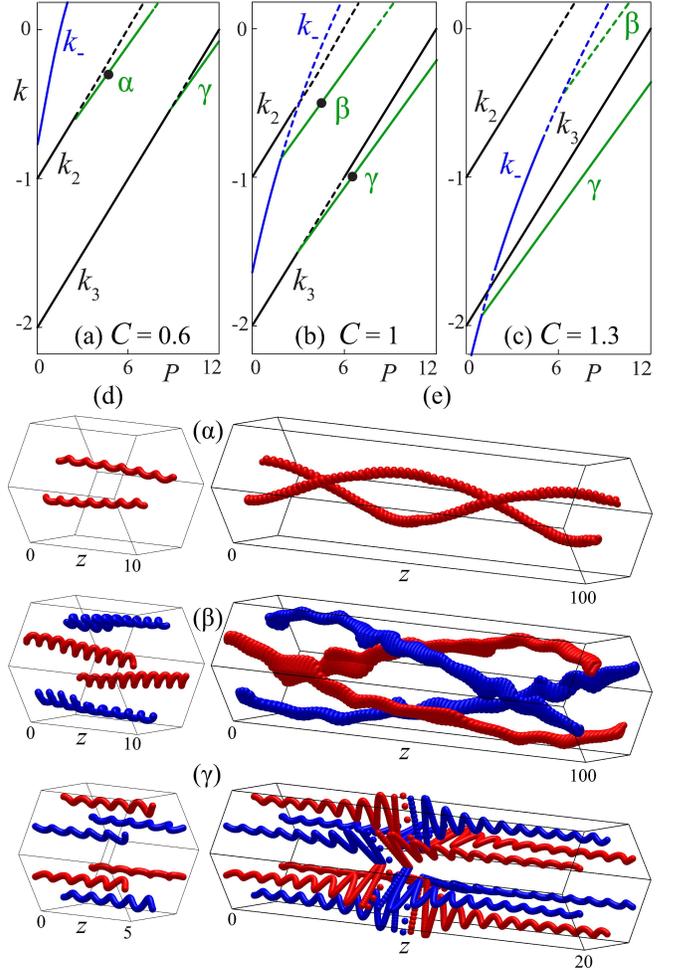}
\caption{(Color online) Existence and stability diagrams for solitons with $\delta_{0,1}=1$ and $C=0.6$ (a), $C=1$ (b), and $C=1.3$ (c). Solid/dashed lines indicate linearly stable/unstable solitons. Blue, black, and green curves correspond to the dark, vortex, and multivortex solitons. Columns (d,e) show trajectories of vortices with charges $\pm 1$ (red and blue lines) for stable perturbed multivortex solitons ($\alpha,\beta,\gamma$) indicated by dots in (a,b). Initial perturbations of solitons are $E_0 = 0.6 A$ for ($\alpha$), $E_0 = 0.8 A$ for ($\beta$), and $|E_n| = 0.8 B$ for ($\gamma$).}
\label{fig:coupling}
\end{figure}

Solutions of particular interest in application to the heptamer in Fig.~\ref{fig:array} are the ones with $N=3|m|$, realizing topological configurations in Figs.~\ref{fig:array}($\alpha,\beta$). Periodicity of Eq.~\eqref{epsi} requires $\psi_{j+3}=\psi_j$ with only three independent phases, $j=1,2,3$. Thus, instead of Eq.~\eqref{epsi} with two complex parameters $b_{\pm m}$, it is convenient to use the identity $\sum e^{i \psi_n}=Nb_N$. We derive from Eqs.~\eqref{solitons}
\begin{eqnarray}
\label{eqn:family}
e^{i \psi_1} + e^{i \psi_2} + e^{i \psi_3}=-C A / B,\label{epsi3}\\
k = \delta_1 B^2-1 = \delta_0 A^2 - C^2 |m|.\nonumber
\end{eqnarray}
The real and imaginary parts of Eq.~\eqref{epsi3} form two equations for three variables $\psi_{1,2,3}$. Therefore, one of three phases is a {\em free parameter}, leaving the total power and the Hamiltonian invariant. The total charge and positions of vortices within the hexagon in Figs.~\ref{fig:array}($\alpha,\beta$) depend on this free parameter as they trace a circle for $CA/B < 1$ with net charge $\pm 2$, while for $CA/B > 1$ they trace a hexagon with zero net charge. We demonstrate in Fig.~\ref{fig:coupling} that such degeneracy of stationary solutions leads to their nontrivial dynamics with spiraling vortices.

In Fig.~\ref{fig:coupling}(a-c) we present existence and stability diagrams for constant-amplitude solitons in the heptamer. At small coupling, such as $C=0.6$ in Fig.~\ref{fig:coupling}(a), the multivortex solitons ($\alpha,\gamma$) bifurcate from vortices with $|m|=2,3$, respectively. As the coupling is gradually increased, the collision of the linear modes marked ($\alpha,\beta$) in Fig.~\ref{fig:array}(b) occurs, and the dark, vortex, and multivortex solitons bifurcate ($P>0$) from this point (not shown). For stronger coupling, such as $C=1$ in Fig.~\ref{fig:coupling}(b), the multivortex soliton ($\beta$) with $m=0$ bifurcates from the dark soliton with $k_-$. With further increase to $C=1.3$ in Fig.~\ref{fig:coupling}(c) the bifurcation for ($\gamma$) soliton changes the ``parent'' branch from $k_2$ in (a,b) to $k_-$ in (c).

\begin{figure}
\centering\includegraphics[width=0.8\columnwidth]{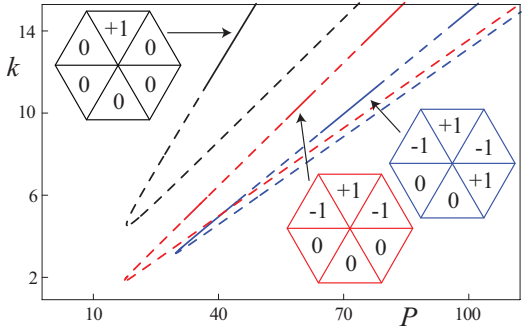}
\caption{(Color online) Examples of stable (solid lines) and unstable (dashed lines) asymmetric multivortex solitons for $N=6$, $\delta_{0,1}=1$, and $C=1$. Each color corresponds to a distinct vortex configuration shown in the insets.}
\label{fig:bifurcations}
\end{figure}

We test the stability of three different solitons, marked by dots in Fig.~\ref{fig:coupling}(a,b), by numerical simulations of initially perturbed stationary solutions. The results are presented in Fig.~\ref{fig:coupling}(d,e) as the trajectories of vortices, recovered from numerical solutions to Eq.~\eqref{DNLS} as described in Ref.~\cite{PRA11}. At small propagation distances in the column Fig.~\ref{fig:coupling}(d) we observe small-scale ``vibrations'' of vortices, with the frequencies given by the eigenvalues of internal modes, as expected for linearly stable solitons. However, at large propagation scales in the column Fig.~\ref{fig:coupling}(e) we observe formation of adiabatically slow superstructures, in particular with vortices spiraling in ($\alpha,\beta$) and moving between triangular cells, thus realizing {\em cyclic charge-flipping}. In contrast, the vortices in the excited multivortex soliton ($\gamma$) oscillate in radial direction and flip their charges simultaneously, via an avoided crossing of vortex lines, similar to the vortex breathers in Ref.~\cite{PRA11}.

We found that the parent branch always loses stability at the multivortex bifurcation point. Thus the appearance of multiple regions of stability and instability in Fig.~\ref{fig:coupling}(a-c) suggests the existence of stable {\em asymmetric} multivortex solitons. Several examples of such solitons, preserving one axis of symmetry, are obtained numerically from Eqs.~\eqref{stat} and presented in Fig.~\ref{fig:bifurcations}. The stability of these stationary states is determined by their global symmetries~\cite{super}, for example the soliton with a single vortex within hexagon [Fig.~\ref{fig:bifurcations}, black lines] has a stability window, while its symmetric counterpart, with the vortex located on the central site, is always unstable~\cite{PRA11}. We also find completely asymmetric stationary states, but in general they appear to be unstable.

In conclusion, we have introduced multivortex solitons in an oligomer cluster of coupled nonlinear oscillators. Such linear and nonlinear states appear at the intersection of vortex and vortex-free modes. The perturbation of stable multivortex solitons leads to the excitation of breather-like states with nontrivial vortex dynamics, including periodic charge flipping and spiraling.

We note that the system considered here resembles the heptamer clusters of plasmonic nanoparticles~\cite{mirin2009,hentschel2010,lassiter2010}. Plasmonic nanoclusters exhibit sharp Fano resonances~\cite{miro2010} in their transmission spectrum, as the consequence of interference between dark and bright modes~\cite{mirin2009} introduced~\cite{hentschel2010} and tuned~\cite{lassiter2010} by the coupling $C$ with the central site. Our results suggest that the collision of plasmonic modes can lead to the excitation of multiple persistent currents in metallic nanoclusters, with additional Fano-like resonances corresponding to different multi-vortex configurations.

This work is supported by the Australian Research Council.

\end{document}